\documentclass[conference]{IEEEtran}
\IEEEoverridecommandlockouts
\usepackage{cite}
\usepackage{amsmath,amssymb,amsfonts}
\usepackage{algorithmic}
\usepackage{graphicx}
\usepackage{textcomp}
\usepackage{xcolor}
\usepackage{mathtools}
\usepackage{fancyhdr}
 \usepackage{multirow}
 \usepackage[a-1b]{pdfx}

\def\BibTeX{{\rm B\kern-.05em{\sc i\kern-.025em b}\kern-.08em
    T\kern-.1667em\lower.7ex\hbox{E}\kern-.125emX}}
\usepackage{nopageno}
\begin{document}

\pagestyle{empty}

\title{A Fair Federated Learning Framework for Collaborative Network Traffic Prediction and Resource Allocation}

\author{\IEEEauthorblockN{Saroj Kumar Panda\IEEEauthorrefmark{1},
Tania Panayiotou\IEEEauthorrefmark{2}, 
Georgios Ellinas\IEEEauthorrefmark{2} and
Sadananda Behera\IEEEauthorrefmark{1}}\\
\IEEEauthorblockA{\IEEEauthorrefmark{1}Dept. of Electronics and Communication Eng., National Institute of Technology Rourkela, Odisha, India\\
\IEEEauthorrefmark{2}KIOS CoE and Dept. of Electrical and Computer Eng., University of Cyprus, Nicosia, Cyprus}}

\maketitle

\thispagestyle{empty}

\begin{abstract}
In the beyond 5G era, AI/ML empowered real-world digital twins (DTs) will enable diverse network operators to collaboratively optimize their networks, ultimately improving end-user experience. Although centralized AI-based learning techniques have been shown to achieve significant network traffic accuracy, resulting in efficient network operations, they require sharing of sensitive data among operators, leading to privacy and security concerns. Distributed learning, and specifically federated learning (FL), that keeps data isolated at local clients, has emerged as an effective and promising solution for mitigating such concerns. Federated learning poses, however, new challenges in ensuring fairness both in terms of collaborative training contributions from heterogeneous data and in mitigating bias in model predictions with respect to sensitive attributes. To address these challenges, a fair FL framework is proposed for collaborative network traffic prediction and resource allocation. To demonstrate the effectiveness of the proposed approach, non-iid and imbalanced federated datasets based on real-word traffic traces are utilized for an elastic optical network. The assumption is that different optical nodes may be managed by different operators.
Fairness is evaluated according to the coefficient of variations measure in terms of accuracy across the operators and in terms of quality-of-service across the connections (i.e., reflecting end-user experience). It is shown that fair traffic prediction across the operators result in fairer resource allocations across the connections. 
\end{abstract}


\section{Introduction}
With the advent of new technologies such as 6G networks, the Internet of Things (IoT), and applications such as virtual reality, smart health, and smart cities, the demand for network connectivity continues to surge. Network service providers (NSPs) face the challenge of upgrading their infrastructure to accommodate the ever-growing traffic needs, while simultaneously minimizing capital and operational expenditures. In this regard, provisioning of traffic-driven services with the use of advanced machine learning (ML) frameworks has been shown to be essential in effectively planning and operating networks with significant spectrum savings, reduced energy consumption, and reduced capital and operational costs~\cite{panayiotou2023survey}. 

Furthermore, digital twins (DTs) technology has recently emerged as a new paradigm for the effective operation of next-generation networks~\cite{10103508}. A digital twin is, in essence, a real-time representation (i.e., virtual replica) of physical entities in the digital world and has been successfully applied in various fields such as Industry 4.0, manufacturing, energy, and healthcare, while the effort is to adopt this technology for telecommunication networks (e.g., 5G/6G and optical networks) ~\cite{9429703,9374645,RAMU2022103663,9356524, 9748222}. In telecommunication networks, which this work focuses on, DTs, especially when empowered by AI/ML, enable realistically simulating the physical network, digitizing associated events in real-time, ultimately providing preventive network maintenance, quality-of-service (QoS) control, network planning, and proactive network (re)optimization, with ML accurately modeling the expected behavior of end-users.  

Amongst the major challenges in DTs is how to efficiently achieve accurate network traffic prediction (that plays a pivotal role in optimizing network planning, resource allocation, and decision-making by capitalizing on the predictability inherent in traffic patterns) and realize the sharing of network resources among various operators, while at the same time enhancing the privacy and security of each operator's data. The integration of edge computing technologies offers a promising solution to these challenges, particularly by enabling intelligence at the network's edge \cite{GUPTA2021103521}.

A cornerstone of edge intelligence is distributed ML. In particular, federated learning~\cite{bonawitz2019towards,mcmahan2017communication,konevcny2016federated}, which allows learning between distributed clients while keeping data local~\cite{li2020federated}, is a technology that could be integrated with DTs to enhance privacy protection and data security~\cite{sun2020adaptivefederatedlearningdigital}. Specifically, in FL, distributed clients train models over their local data, and only model updates are shared with a central server for aggregation, thereby avoiding the need for transferring raw data. This distributed learning approach not only mitigates privacy and security concerns but also reduces communication costs~\cite{mcmahan2017communication}, making it an ideal framework for applications requiring real-time, privacy-sensitive data processing.


\subsection{Related Work}
In the context of network traffic prediction, FL allows distributed training between multiple clients, with each client having access to the data of a different operator~\cite{sepasgozar2022fed,li2024core}. This enables the creation of a centralized predictive traffic demand matrix (e.g., formed in a cloud orchestrator), enabling the proactive (re)optimization of shared network resources~\cite{9748600} with increased privacy and security. In fact, distributed learning~\cite{Chen:19} and traditional FL, is previously considered in~\cite{behera2024federated,drainakis2023centralized,9306745,9748424,asad2021federated,9605846}, where the ability of FL to achieve performance accuracies comparable to the centralized learning paradigm is demonstrated. However, in those works, federated datasets are assumed to be independent and identically distributed (iid), largely ignoring the fact that the various traffic sources may be heterogeneous.

In the presence of heterogeneous traffic (i.e., imbalanced and non-iid datasets), traditional FL fails to achieve sufficiently high accuracy across all clients, since it focuses on minimizing a global (aggregated) loss function across participating clients, which may lead to a disproportionate advantage (or disadvantage) for some of the clients. In other words, traditional FL does not inherently account for achieving fair training accuracy across the different clients; rather, its target is to maximize the aggregated accuracy over all clients.


To address this challenge, a fair FL approach is considered that introduces a mechanism to enhance fairness by focusing on the performance of clients with higher empirical losses. In general, fairness-aware ML has recently attracted growing attention~\cite{ezzeldin2023fairfed,salazar2023fair,li2019fair}. Common approaches include pre-processing the data to remove sensitive information or post-processing the model by adjusting the prediction thresholds after training~\cite{hardt2016equality,calmon2017optimized}. Another set of works focuses on optimizing the learning objective under fairness constraints during training \cite{agarwal2018reductions,hashimoto2018fairness,cotter2019optimization}. 


\subsection{Contribution}
In this work the $q$-Fair FL ($q$-FFL) scheme~\cite{li2019fair} is utilized, in which fairness is defined as the uniformity of prediction accuracy across all clients and it can be controlled by appropriately tuning a single hyperparameter ($q$). Without loss of generality, an optical network infrastructure is assumed, where each client represents a different operator (i.e., an SDN controller is assumed over the optical node/s~\cite{9748222} managed by the same operator). First, the $q$-FFL framework is examined, for various $q$ values, over non-iid and imbalanced datasets generated according to real-word traffic traces. To directly interpret fairness with respect to the achievable accuracy across the operators, the coefficient of variations (CV) metric is employed~\cite{jain1984quantitative}. Results indicate that as $q$ increases, fairness in accuracy across the operators improves. Importantly, to examine the impact of the various $q$-fair models on the optical connections, the test traffic predictions are used to guide the spectrum allocation (SA) decisions. The CV metric is utilized to measure the fairness with respect to the achievable QoS across all network connections, demonstrating that fairer traffic prediction models lead to fairer QoS-based resource allocations. To the best of our knowledge, this is the first time that fair FL is examined for network traffic prediction, demonstrating also the impact of fairer predictions on the resource usage within optical networks (leading to a fairer over-under-provisioning trade-off). Overall, this is a framework that is well-suited for diverse and privacy-sensitive network environments, and can thus be combined with DTs for effective traffic prediction and resource allocation, while enhancing privacy and security.       

\section{Fair Federated Traffic Modeling}
The preliminaries regarding the $q$-FFL approach are initially discussed, followed by the description of the $q$-FFL algorithm developed and the formulation of the federated network traffic prediction problem.


\subsection{Fair Federated Learning Preliminaries}
In the traditional FL approach, the objective is to 
minimize a global objective function $f(w)$:  


\begin{align}
f(w) = \sum_{k=1}^{m} p_{k}F_{k}(w)
\label{eq1}
\end{align}

\noindent where $m$ is the number of clients, $F_{k}$ is the loss function at the $k$th client, and $p_{k}$=$\frac{n_{k}}{n}$, with $n_{k}$ denoting the number of samples available at client $k$ and $n$=$\sum_{k} n_{k}$ denoting the total number of samples. On the contrary, $q$-FFL introduces a new optimization function inspired by the $\alpha$-fairness scheme~\cite{5461911}, that minimizes the following global objective function ($f_q(w)$): 


\begin{align}
f_q(w) = \sum_{k=1}^{m} \frac{p_k }{q+1} F_k^{q+1}(w),
\label{qffl}
\end{align}

\noindent where $q$ is a scalar parameter that similarly to the inequality aversion parameter $\alpha$, controls fairness across the clients. Specifically, when $q=0$, then $q$-FFL reduces to the traditional FL objective (Eq.~\ref{eq1}) to minimize the global training loss $f_q$, or equivalently, to maximize efficiency in global model accuracy. As $q$ increases, more uniformity is imposed on the training loss, improving the fairness across the clients. Finally, setting $q$ to a value large enough reduces to the classical min-max fairness~\cite{mohri2019agnostic} that minimizes the maximum loss encountered at a client. However, by increasing fairness the global model accuracy is likely to decrease~\cite{Bertsimas11}, and the same holds for the maximum accuracy that a client may achieve. As such, it is important that $q$ is properly tuned given the application at hand.   


\subsection{$q$-FFL Algorithm}
\label{fair_fl}
In a nutshell, during training, the loss function gives more weight to clients with higher losses, aiming to reduce performance disparities across the clients. Specifically, each client, trains a local model on its local traffic traces and computes its own loss function $F_{k}$, raised to the power of $q$, so that clients with higher losses have a proportionally greater impact.

Then, a cloud orchestrator aggregates the weighted updates received from all clients using a fairness-aware approach to update the global model. In this approach, as the updates from clients with higher losses are weighted more heavily (i.e., more consideration is given to clients with higher losses), the global model is updated in a way that improves fairness across the clients (i.e., across the operators). Once the global model is updated, it is sent back to all participating clients. The process of local training, aggregation, and global model updates, is repeated over several communication rounds until the model converges or a predefined stopping criterion is met. 
Local training is carried out using stochastic gradient descent (SGD), while aggregation of the weighted updates from all clients is carried out following a federated averaging approach~\cite{li2019fair}. 
For both global and local models, long-short-term memory (LSTM) neural networks are considered~\cite{greff2016lstm}, as, in general, deep learning algorithms with recurrent units (e.g., LSTMs) have shown to be capable of capturing the non-linear long-term dependencies in network traffic sequences~\cite{panayiotou2023survey}. Even though other models could also be considered (e.g., Transformers~\cite{NIPS2017_3f5ee243}), their examination and a comparison between different models is planned as future work.  

\subsection{Traffic Prediction Modeling and Formulation}
In this work, for simplicity, it is assumed that an SDN controller stores and analyzes the traffic data of one network node or equivalently, that each participating operator manages a single node. Hence, for each client $k$, a different federated dataset $D_k=\{{\bf x}^{(k)'}_t, {\bf x}^{(k)}_t \}_{t=1}^{n_k}$ is created, where ${\bf x}^{(k)'}_t$ is a traffic sequence that includes the past and present traffic observations:


\begin{equation*}
{\bf x}^{(k)'}_t=[x^{(k)}_{t-\kappa},..,x^{(k)}_{t-1}, x^{(k)}_{t}],
\end{equation*}


\noindent and ${\bf x}^{(k)}_t$ is the next traffic observation that the federated model is trained to predict:


\begin{equation*} 
{\bf  x}^{(k)}_t=x^{(k)}_{t+1},
\end{equation*} 

\noindent In this setting, $\kappa$ is the number of past observations, $t$ refers to the current time instant, and past and present traffic data are collected at regular intervals of $\tau$ time units (e.g., hourly). Thus, traffic observations ${\bf x}^{(k)'}_t$ and ${\bf x}^{(k)}_t$ are sequential in time and are spaced $\tau$ units apart.

The $q$-FFL model is trained on datasets $D_k$. 
After model training, each federated LSTM model can predict, at any time instant $*$, the future traffic value ${\bf \hat x}^{(k)}_*$, given the traffic sequence ${\bf x}^{(k)'}_*$ that is locally generated at each network node $k$. This information is communicated to the cloud orchestrator to be used for proactive resource allocation and network (re)optimization.  

\section{Federated Dataset Generation}
\label{dataset}
The performance of the FL traffic prediction framework is evaluated according to the real-world traffic dataset obtained from the $12$-node, $30$-link Abilene backbone network (http://sndlib.zib.de/home.action). This dataset provides bit-rate information (in Gbps) for every node pair in the form of traffic demand matrices, given every $5$-minutes for a $6$-month period. In this work, a dataset $D_k$ is created for each node $k$ to represent the aggregated bit-rate to this node in $5$-minute intervals. Given the aggregated bit rates, traffic patterns for each sequence in $D_k$ (i.e., $[{x}^{(k)}_{t-\kappa}, \cdots, {x}^{(k)}_{t-1},{x}^{(k)}_t, {x}^{(k)}_{t+1}]$) are then created with $\kappa=70$, following the sliding window approach.

In total, $m=5$ (i.e., network nodes) are considered and for each federated dataset $D_k$ a different size is considered to additionally examine the $q$-FFL performance over imbalanced datasets. Specifically, $D_1=3000$, $D_2=2000$, $D_3=8000$, $D_4=5000$, and $D_5=7500$, with these patterns being sequential in time. Further, to introduce heterogeneity into the federated datasets (that is, to form non-iid datasets), noise is sampled and infused into the traffic traces of some of the datasets. Noise values are drawn from various distributions with different parameters, and these noise values are added to the traffic traces of the datasets considered. Specifically, traffic traces in $D_1$ are infused with noise values drawn from the Gaussian distribution ($\mu$=$10$, $std$=$2$), while traffic traces in $D_2, D_3$, and $D_4$ are infused with noise values drawn from the log-normal distribution ($\mu$=$1$, $std$=$0.5$), exponential distribution ($\lambda$=$2$), and Gamma distribution ($\alpha$=$1$, $\beta$=$3$), respectively. It is worth mentioning that $q$-FFL was trained on the imbalanced federated datasets prior to noise infusion, but it was shown that those datasets were iid, since all the global and local losses across the datasets were invariant over all the $q$ values examined.      

\section{Fair FL Training and Evaluation}
To set up the $q$-FFL framework, Python version 3.9.0 along with TensorFlow 2.10.1 is utilized on a computing system equipped with an Intel(R) Core(TM) i5-10300H CPU running at 2.50 GHz and 16 GB of RAM. For data preprocessing, \emph{StandardScaler} is employed to normalize the datasets. Each dataset $D_k$ is then partitioned such that the last $100$ sequences are retained for testing and from the remaining sequences $80\%$ are used for training while the other $20\%$ are used for validation. 
For local model training, the following parameters are considered: batch size=$256$, learning rate=$10^{-4}$, using a sequential model with $2$ LSTM hidden layers trained to optimize the mean squared error (MSE) loss function (i.e., the $F_k$). For global model optimization, $100$ iterations are performed.

\begin{figure*}[h]
\centering
\includegraphics[scale=0.6]{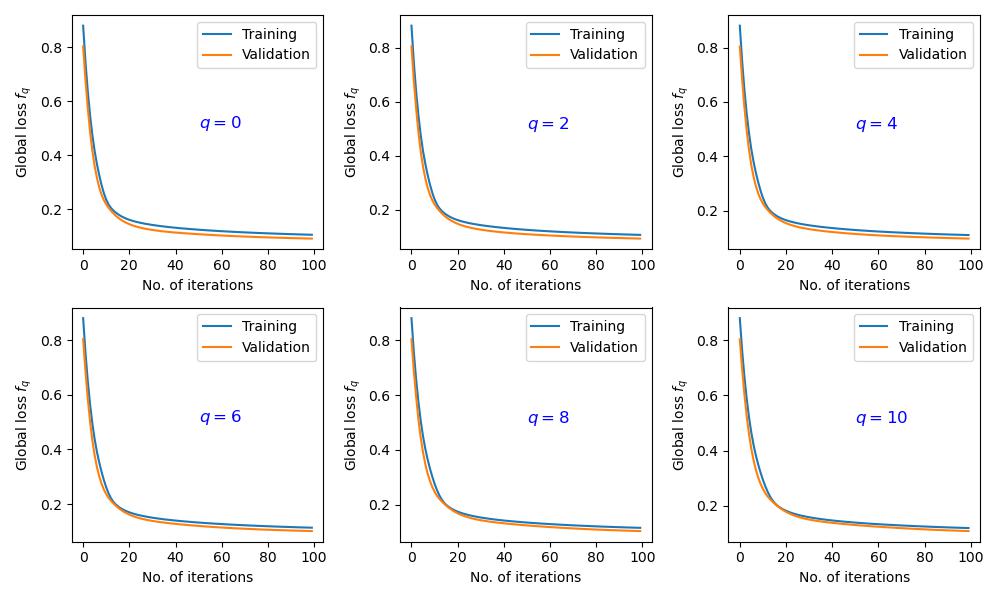} 
\caption{The global loss $f_q$ across all clients for different $q$ values.}
\label{learning_curves}
\end{figure*}

Figure~\ref{learning_curves} presents the training and validation global loss (i.e., $f_q$) over the number of training iterations for the various $q$ values considered. It can be observed that both training and validation losses converge to an MSE close to zero for all federated datasets and for all $q$ values examined. Note that while the losses for intermediate $q$ values (e.g., $q$ = 1, 3, 5, 7, 9) also exhibit similar convergence behavior, they are omitted from the figure due to space constraints.

For a better insight into the derived $q$-fair models, Table~\ref{MSE_COV} presents for each client the test loss $F_k$ and the global test loss $f_q$, calculated over the test traffic traces of each $D_k$ (that is, the last $100$ sequences in each federated dataset). According to Table~\ref{MSE_COV}, global test loss $f_q$ remains relatively consistent with increasing $q$, an indicator that in the problem setting examined, fairer traffic prediction models do not affect efficiency in accuracy. 

\begin{table}[h]
\centering
\caption{Local loss $F_k$ and global loss $f_q$ for different $q$ values.}
\label{MSE_COV}
\begin{tabular}{|l|l|l|l|l|l|l|}
\hline
q  & $F_1$ & $F_2$ & $F_3$ & $F_4$ & $F_5$ & \multicolumn{1}{c|}{\begin{tabular}[c]{@{}c@{}} $f_q$\end{tabular}}   \\ \hline
0  & 0.2776   & 0.0558   & 0.0950   & 0.1746   & 0.1889   & 0.1584 \\ \hline
2  & 0.2443   & 0.0603   & 0.0922   & 0.1798   & 0.1879   & 0.1529  \\ \hline
4  & 0.2324   & 0.0635   & 0.0918   & 0.1845   & 0.1903   & 0.1525  \\ \hline
6  & 0.2290   & 0.0654   & 0.0924   & 0.1883   & 0.1935   & 0.1537  \\ \hline
8  & 0.2293   & 0.0666   & 0.0937   & 0.1916   & 0.1966   & 0.1555 \\ \hline
10 & 0.2313   & 0.0673   & 0.0952   & 0.1943   & 0.1991   & 0.1575 \\ \hline
\end{tabular}
\end{table}

Fairness is evaluated according to the CV measure to capture the variability in accuracy across the clients. In general, lower CV values indicate fairer solutions, with the zero value indicating that the fairest solution is derived (i.e., for the FFL problem this would indicate a uniform distribution in accuracy achieved across the clients). Specifically, in this work, to evaluate the $q$-fair models, the CV of achievable accuracy is measured for each $q$ according to:

\begin{equation}
CV^{loss}_q=100 \times \sqrt{\frac{m^2}{m-1} \frac{\sum_{k}^m (F_{k} -\frac{1}{m}\sum_{k}^m  F_{k})^2}{(\sum_{k}^m  F_{k})^2}},
\label{fairness_metric}
\end{equation}

\begin{figure}[h]
\centering
\includegraphics[scale=0.3]{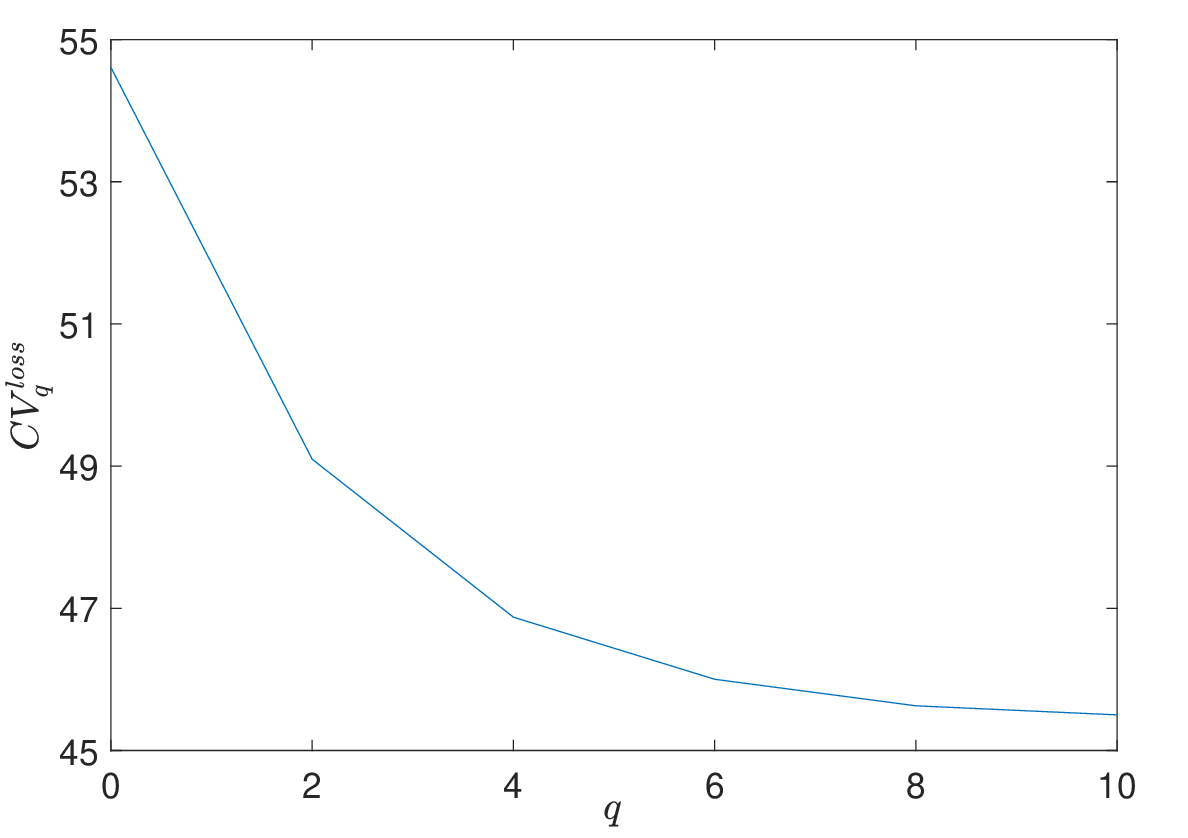} 	
\caption{$CV$ of accuracy across all clients for various $q$ values.}
\label{cov_mse}
\end{figure}

Figure~\ref{cov_mse} illustrates $CV^{loss}_q$ for the various $q$ values considered. Clearly, as $q$ increases, the CV consistently decreases, indicating a reduction in the variability across clients and leading to a more uniform, and thus fairer, distribution in accuracy across the operators. Even though higher values of $q$ (i.e., $q>10$) were also examined, fairness did not significantly improve, which is an indicator that a $q$-fair model has been reached, that is approaching min-max fairness. Overall, this min-max fair model (i.e., for $q=10$) improved fairness by approximately $16\%$ when compared to the least fair model (i.e., $q=0$), significantly improving equity in the accuracy across the clients. 

\section{Fair Traffic-Driven Resource Allocation and Evaluation}
In this section the aim is to examine the impact of fairer traffic predictions on the resource allocation decisions and consequently on the QoS of the network connections established. In general, when fairness in accuracy across the federated datasets is not considered, certain operators may be favored by achieving a higher prediction accuracy than others, with this situation possibly leading to an unfair allocation of the shared network resources. Specifically, the favored operators, and consequently the connections they manage, may enjoy a better QoS than the operators achieving a lower prediction accuracy. To examine the impact of fairer traffic-driven resource allocation decisions on the network connections an elastic optical network (EON) is assumed, where flexibility and fine granularity in spectrum allocation (SA) is possible, rendering accuracy in traffic prediction critical in avoiding inefficiencies in network performance (i.e., in connection over- and under-provisioning). 

The assumption is that the traffic predictions from all operators are communicated periodically to the central orchestrator to form the traffic prediction matrix. The predictive matrix is then used to proactively solve the routing and spectrum allocation (RSA) problem. The RSA problem is solved by considering the shortest path routing algorithm and the fist-fit spectrum allocation (SA) scheme. For the routing decisions, the Abilene backbone network topology is considered (described in Section~\ref{dataset}), assuming, for simplicity, infinite spectrum capacity for each network link. 

In particular, for each client/operator $k$, managing the source node $k$ in the Abilene network, a destination node $d$ is randomly chosen to form the $k-d$ pair and solve the routing problem. Given the calculated route, the $q$-fair predicted data rates (i.e., ${\bf \hat x_{*}}^{(k)}$ in Gbps) are converted into the required number of spectrum slots (i.e., ${\bf \hat y_{*}}^{(k)}$) using binary phase shift keying (BPSK) modulation (worst-case scenario). Assuming that BPSK modulation allows for a data transmission rate of $10$ Gbps per $12.5$ GHz frequency slot in an EON~\cite{8681101}, the required number of spectrum slots can be calculated as ${\bf  \hat y_{*}}^{(k)}=\lceil {\bf \hat x_{*}}^{(k)}/10 \rceil$. The RSA problem solved for a $k-d$ pair, then forms Connection $k$.

Subsequently, under-provisioning and over-provisioning are calculated, as measures of QoS, by comparing, for each connection $k$, the $q$-fair predicted number of spectrum slots with the actual number of spectrum slots (obtained from the ground-truths in the test dataset). Specifically, over-provisioning is given by $o_k=\sum_* {\bf \hat y_{*}}^{(k)}-{\bf y_{*}}^{(k)}$ for all ${\bf \hat y_{*}}^{(k)}>{\bf y_{*}}^{(k)}$, and under-provisioning is given by $u_k=\sum_* {\bf \hat y_{*}}^{(k)}-{\bf y_{*}}^{(k)}$ for all ${\bf \hat y_{*}}^{(k)}<{\bf y_{*}}^{(k)}$, where ${\bf y_{*}}^{(k)}$ is the true number of requested spectrum slots. 

Both under- and over-provisioning pose challenges to efficient network operations; under-provisioning can lead to QoS degradation, while over-provisioning wastes spectrum resources. Therefore, achieving a balance between fairness and prediction accuracy is essential for efficient resource management in EONs. Table~\ref{up_op} summarizes the over- and under-provisioning results obtained according to various $q$ values (different fairness levels), providing also the average over- and under-provisioning results, obtained according to $\hat{o}=1/m\sum_k^m o_k$ and  $\hat{u}=1/m\sum_k^m u_k$, respectively. These results can help in quantifying how 
fairness affects resource allocation decisions. Specifically, to directly interpret fairness in resource allocation, the CV of QoS across the connections is calculated as:

\begin{equation}
CV^{QoS}_q=100 \times \sqrt{\frac{1}{2m-1} \frac{\sum_{k}^m (u_{k}  -\hat{Q})^2 + \sum_{k}^m (o_{k}  -\hat{Q})^2}{\hat{Q}^2}},
\label{fairness_QoS_metric}
\end{equation}
where $\hat{Q}=\frac{1}{2m} \sum_{k}^m  (u_{k}+o_{k})$ (i.e., the mean over- and under-provisioning over all connections).

\begin{figure}[h!]
\hspace{-.1 in}
\centering
\includegraphics[scale=0.3]{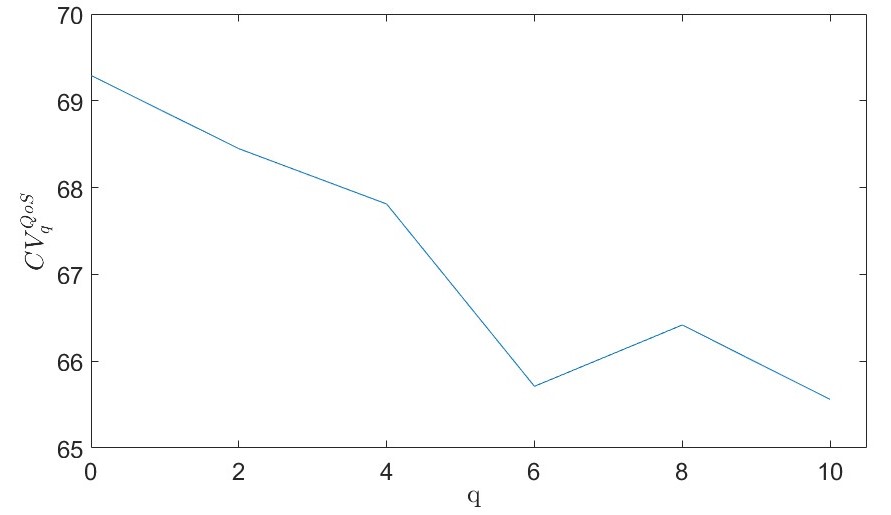} 
\caption{CV of QoS across all the connections for various $q$ values.}
\label{cov_qos}
\end{figure}

The $CV^{QoS}_q$ metric over the different $q$ values is illustrated in Fig.~\ref{cov_qos}, indicating that fairer federated models result in fairer resource allocations. Overall, an improvement up to $6\%$ between the least fair ($q$=$0$) and min-max fair $q$=$10$ solutions is observed in $CV^{QoS}_q$. This improvement is expected to increase as the number of federated datasets increases (i.e., with increasing heterogeneity). However, a larger number of federated datasets is not examined in this work due to the high processing time required for training the $q$-FFL (several training hours required for each $q$ considering $5$ federated datasets). A slight fluctuation that is observed in Fig.~\ref{cov_qos} is mainly due to the conversion between Gbps to spectrum slots that may distort the impact of fairer federated traffic predictions on the spectrum allocation. The trend, however, in Fig.~\ref{cov_qos} is decreasing as $q$ increases (i.e., in the long run fairer resource allocations are derived). 

\begin{table*}[h!]
\centering
\caption{Under-provisioning ($u_k$) and over-provisioning ($o_k$) for different $q$ values.}
\label{up_op}
\begin{tabular}{|c|cc|cc|cc|cc|cc|c|c|}
\hline
\multirow{2}{*}{q} & \multicolumn{2}{c|}{Connection 1} & \multicolumn{2}{c|}{Connection 2} & \multicolumn{2}{c|}{Connection 3} & \multicolumn{2}{c|}{Connection 4} & \multicolumn{2}{c|}{Connection 5}  & \multirow{2}{*}{\begin{tabular}[c]{@{}c@{}} $\hat{u}$\end{tabular}} & \multirow{2}{*}{\begin{tabular}[c]{@{}c@{}} $\hat{o}$\end{tabular}} \\ \cline{2-11}
& \multicolumn{1}{c|}{$u_k$}  & $o_k$ & \multicolumn{1}{c|}{$u_k$}  & $o_k$ & \multicolumn{1}{c|}{$u_k$}  & $o_k$ & \multicolumn{1}{c|}{$u_k$}  & $o_k$ & \multicolumn{1}{c|}{$u_k$}  & $u_k$  &  & \\ \hline
0 & \multicolumn{1}{c|}{104} & 75 & \multicolumn{1}{c|}{54}  & 39 & \multicolumn{1}{c|}{12}  & 36 & \multicolumn{1}{c|}{53}  & 47 & \multicolumn{1}{c|}{89}  & 182 & 62.4 & 75.8 \\ \hline
2 & \multicolumn{1}{c|}{107} & 68 & \multicolumn{1}{c|}{59}  & 41 & \multicolumn{1}{c|}{13}  & 36 & \multicolumn{1}{c|}{54}  & 46 & \multicolumn{1}{c|}{98}  & 182 & 66.2 & 74.6 \\ \hline
4 & \multicolumn{1}{c|}{107} & 66 & \multicolumn{1}{c|}{64}  & 40 & \multicolumn{1}{c|}{13}  & 37 & \multicolumn{1}{c|}{56}  & 47 & \multicolumn{1}{c|}{103} & 183 & 68.6  & 74.6 \\ \hline
6 & \multicolumn{1}{c|}{107} & 65 & \multicolumn{1}{c|}{67}  & 40 & \multicolumn{1}{c|}{13}  & 38 & \multicolumn{1}{c|}{57}  & 47 & \multicolumn{1}{c|}{105} & 177 & 69.8 & 73.4 \\ \hline
8 & \multicolumn{1}{c|}{109} & 66 & \multicolumn{1}{c|}{69}  & 38 & \multicolumn{1}{c|}{13}  & 38 & \multicolumn{1}{c|}{56}  & 49 & \multicolumn{1}{c|}{106} & 180 & 70.6 & 74.2 \\ \hline
10 & \multicolumn{1}{c|}{108} & 66 & \multicolumn{1}{c|}{70}  & 39 & \multicolumn{1}{c|}{13}  & 38 & \multicolumn{1}{c|}{60}  & 49 & \multicolumn{1}{c|}{107} & 180 & 71.6  & 74.4 \\ \hline
\end{tabular}
\end{table*}

Additionally, the fairness in terms of resource allocation across the average over- and under-provisioning metris is calculated for the various $q$ values. This is achieved by calculating the CV of $o/u$ as:

\begin{equation}
CV^{o/u}_q=100 \times \sqrt{\frac{ ((\hat{u}+ \hat{o}) -\frac{1}{2}(\hat{u}+ \hat{o}))^2}{ (\hat{u}+ \hat{o})^2/4}}. 
\label{fairness_ou_metric}
\end{equation}

\begin{figure}[h!]
\centering
\includegraphics[scale=0.3]{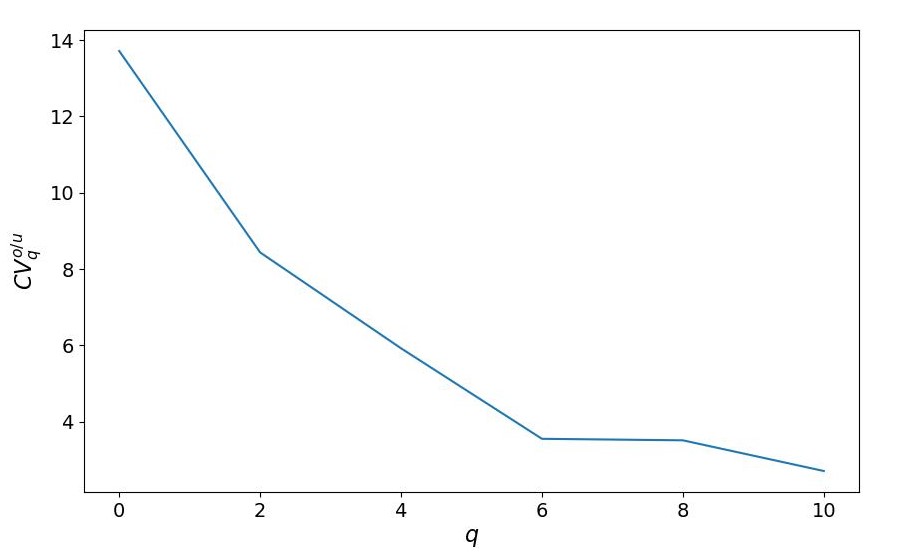} 
\caption{CV across over- and under-provisioning for various $q$ values.}
\label{cov_ou}
\end{figure}

The $CV^{o/u}_q$ metric over the different $q$ values is illustrated in Fig.~\ref{cov_ou}, indicating that fairer federated models result in fairer resource allocation across the average over- and under-provisioning results. Overall, an improvement up to $80\%$ between the least fair ($q=0$) and the min-max fair ($q=10$) solutions is observed in $CV^{o/u}_q$. This result indicates that fairer federated models across the operators better balance average over- and under-provisioning (i.e., approaching equity as $q$ increases). This is also obvious by observing Table~\ref{up_op}, and specifically, the $\hat{u}$ and $\hat{o}$ values as $q$ increases. For the min-max fair solution (i.e., $q=10$) it is obvious that $\hat{u}$ and $\hat{o}$ approach equity. This is the outcome of in essence balancing, in a fairer manner, the prediction error around the true traffic values.   

It should be noted that although in this work fairer resource allocations are derived at the expense of under-provisioning (i.e., according to Table~\ref{up_op} under-provisioning is increased for the sake of improving over-provisioning), this is something that greatly depends on the problem setting and the federated datasets; that is, it is possible that different federated datasets may result in an increased over-provisioning for the sake of improving under-provisioning. Further, for the problem setting considered, fairer federated models did not result in a reduction on the global model accuracy. However, this is not always the case; rather, the rule of thumb is that fairer federated models are more likely to lead to a degradation in global accuracy. Hence, a network operator will need to evaluate various federated models over various $q$ values in order to identify such trade-offs and to decide which $q$-fair model best meets its targets (e.g., in terms of QoS offered).

\section{Conclusions}
\label{conclusions}
In this work, a $q$-FFL framework for network traffic prediction and resource allocation in EONs is examined, assuming imbalanced and non-iid federated datasets. First, the capabilities of the $q$-FFL framework are leveraged by finding federated models that are fairer in terms of accuracy across the operators (up to $16\%$ fairer when compared to the traditional FL approach). 
Subsequently, the $q$-fair federated models are used to guide the spectrum allocation decisions in the network. It is shown that fairer federated models result also in fairer resource allocations in the achievable QoS across the connections (up to $6\%$ improvement), an indicator of the importance of considering during training the heterogeneity of federated datasets in traffic-driven resource allocation. Importantly, fairer federated models lead to a fairer over-under-provisioning trade-off, with the min-max solution indicating an improvement of $80\%$. Clearly, this framework can be incorporated in DTs to efficiently realize the sharing of network resources among various operators. Further, as real-word DTs are expected to involve more operators/federated datasets in the future, fairness considerations are expected to become even more critical.     
   
As network operators, contending for the shared network resources, cannot know a-priori which is the most effective $q$ value to consider, this framework must be evaluated over various $q$ values. Developing a fair FL framework that is independent of manually tuning the level of fairness is planned for future work. Furthermore, this framework can be further improved through explainable artificial intelligence (XAI)~\cite{10527146,10619785}. 

\section*{Acknowledgments}
This work has been supported by the European Union’s Horizon 2020 research and innovation programme under grant agreement No. 739551 (KIOS CoE - TEAMING) and from the Republic of Cyprus through the Deputy Ministry of Research, Innovation, and Digital Policy.

\bibliographystyle{IEEEtran}
\bibliography{IEEEabrv,ref_v2.bib}

\end{document}